\documentclass[reprint,amsmath,amssymb,aps,prb,showpacs,floatfix]{revtex4-1}

\usepackage[dvips]{color,graphicx}
\DeclareGraphicsExtensions{.eps, .jpg, .png}

\begin{document}

\title{Quantum Monte Carlo study of the three-dimensional spin-polarized homogeneous electron gas}

\author{G.\ G.\ Spink}
\affiliation{TCM Group, Cavendish Laboratory, University of Cambridge, 
      J.\ J.\ Thomson Avenue, Cambridge CB3 0HE, United Kingdom}

\author{N.\ D.\ Drummond}
\affiliation{Department of Physics, Lancaster University, Lancaster LA1 4YB,
  United Kingdom}

\author{R.\ J.\ Needs}
\affiliation{TCM Group, Cavendish Laboratory, University of Cambridge, 
      J.\ J.\ Thomson Avenue, Cambridge CB3 0HE, United Kingdom}

\date{\today}

\begin{abstract} We have studied the spin-polarized three-dimensional
  homogeneous electron gas using the diffusion quantum Monte Carlo method,
  with trial wave functions including backflow and three-body correlations 
  in the Jastrow factor, and we have used twist averaging to reduce finite-size 
  effects.  Calculations of the pair correlation function,
  including the on-top pair density, as well as the structure factor and the total
  energy, are reported for systems of 118 electrons in the density range
  $r_\text{s}=0.5$--$20$ a.u., and for spin polarizations of 0, 0.34, 0.66, and 1.
  We consider the spin resolution of the pair correlation function and
  structure factor, and the energy of spin polarization.  We show that a control 
  variate method can reduce the variance when twist-averaging, and we have achieved 
  higher accuracy and lower noise than earlier quantum Monte Carlo studies.
 \end{abstract}


\maketitle


\section{Introduction \label{sec:intro}}

The simplicity of the three-dimensional (3D) homogeneous electron gas
(HEG) model,\cite{giuliani} consisting of electrons interacting via
the Coulomb potential in a uniform, positive background for charge
neutrality, allows the study of important features of the
many-electron problem without the complication of a lattice potential.
Early pioneers of quantum mechanics discovered much of the
phenomenology of the HEG, including Wigner's celebrated
insight\cite{Wigner,CepAlder} that the fluid will crystallize at low
density.  More recently, the continuing research effort to understand
the behavior of the HEG is motivated by the model's application in
density functional theory (DFT)\@.

Quantum Monte Carlo (QMC) methods can provide accurate estimates of
the static properties of interacting many-body systems such as the
HEG\@.  The ground state of the 3D HEG for collinear spins is accepted
to be an unpolarized Fermi fluid at high densities.  As the density
decreases, QMC calculations indicate an apparently continuous
transition to a spin-polarized fluid occurs, starting at a density of
about $r_\text{s}=50 \pm 2$ a.u.\cite{Zong} The transition to a Wigner
crystal is calculated to occur at about $r_\text{s}=106 \pm 1$
a.u.\cite{QMC-Wigner} QMC calculations have furnished data used within
DFT, most notably the energies of the HEG calculated by Ceperley and
Alder,\cite{CepAlder} which are used to parameterize the
local-density-approximation exchange-correlation functional.
Semilocal\cite{semilocal} and nonlocal\cite{nonlocal1,nonlocal2}
exchange-correlation functionals may use additional properties of the
HEG such as the pair correlation function (PCF) and, especially, the
on-top pair density.\cite{g(0)-DFT}

The PCF is a measure of the spatial correlations of electron positions
that arise from the Coulomb repulsion and Pauli exclusion.  The
spin-resolved PCF of the HEG, $g_{\alpha\beta}(r)$, is defined such
that $n_{\beta} g_{\alpha\beta}(r) 4\pi r^{2}dr$ is the expected
number of spin-$\beta$ electrons in an infinitesimal shell of radius
$r$ when a spin-$\alpha$ electron is found at the origin and
$n_{\beta}$ is the number density of spin-$\beta$ electrons.  Defining
the spin polarization $\zeta$ as the ratio
$\zeta=(N_{\uparrow}-N_{\downarrow})/(N_{\uparrow}+N_{\downarrow})$,
the total PCF $g$ is a weighted average of the spin-resolved components:
\begin{equation}
  g=\left( \frac{1+\zeta}{2} \right)^2 g_{\uparrow \uparrow} + 
  \left( \frac{1-\zeta}{2} \right)^2 g_{\downarrow \downarrow} +  \\
  \left( \frac{1-\zeta^2}{2} \right) g_{\uparrow \downarrow}.
 \label{eq:weighted-average}
\end{equation}
The on-top pair density $g(0)$ is the value of the PCF at contact.
The region around $r=0$ is, however, the most challenging to sample in
a stochastic simulation, particularly at low densities, as the
electrons are rarely found close to one another.  Toulouse \emph{et
  al.}\cite{TAU}\ have developed a possible solution to this
difficulty, extending the zero-variance zero-bias estimators of
Assaraf and Caffarel.\cite{AC} It has recently been found by
Fantoni,\cite{Fantoni} however, that the zero-bias correction required
in DMC calculations significantly increases the variance.  We
therefore opted to use the traditional histogram estimator of the PCF
in this work; our simulations gathered enough data to ensure good
precision in the reported $g(0)$ values.

In light of the above challenges, while QMC methods have been very successful 
in calculating many quantities of interest in the HEG, such as the energy and the PCF at intermediate 
distances, the short-range behavior of the PCF has usually been obtained by fitting QMC data to some
analytical function, which can be constrained to obey exact results such as
the electron-electron cusp conditions.  Even then, however, there is some
disagreement between the PCFs obtained using different approaches.
Gori-Giorgi and Perdew developed an analytical model based on exact
constraints and QMC energy data.\cite{GP2,GP3} The same authors\cite{GP1}
extended an approach originally due to Overhauser\cite{Overhauser} using a
screened Coulomb potential and two-electron wave functions. An alternative
starting point is afforded by ladder theory, which gave rise to an early and
widely used result for $g(0)$ due to Yasuhara.\cite{Yasuhara} Nagy \emph{et
  al.}\cite{Nagy}\ offer insight into the good agreement between these results.
More recently, Qian\cite{Qian} was able to relax an approximation made in the
earlier ladder theory calculations and obtained a markedly different
result. Although several QMC calculations of PCFs for the unpolarized 3D HEG
have been reported, less attention has been paid to fully and, especially, 
partially spin-polarized systems.\cite{OB94,Pickett,OHB,GSB,Holzmann,Eclr,Fantoni}

In this paper we present extensive results for PCFs, static structure
factors (SSFs), and energies for HEGs containing 118 electrons over the
density range relevant to DFT calculations, $r_\text{s}=0.5$--20 a.u.
We consider four values of the spin polarization,
$\zeta=0$, $40/118$, $78/118$, and $1$, although in the text we refer to the
intermediate polarizations as 0.34 (40/118) and 0.66 (78/118).  We use
Slater-Jastrow wave functions incorporating backflow and three-body
correlations, we investigate twist-averaged boundary conditions and
finite-size effects, and we are able to gather sufficient data to
ensure that the statistical uncertainty is relatively modest. A variance 
reduction method is found to improve significantly the precision of 
twist-averaged data without introducing bias.  We show 
the spin resolution of the PCFs and the SSFs, and we consider in
particular the on-top pair density and the energy of spin polarization as
functions of $r_\text{s}$ and $\zeta$.  We have performed spline fits
to our PCF data, which appear to be more accurate than polynomial fits. 
A short Fortran 90 program is available to evaluate the spline fits to our
PCF data.\cite{supplemental}

Our paper is organized as follows.  In Sec.\ \ref{subsec:methods} we
discuss the QMC methods used in our calculations.  Section
\ref{sec:results} describes our PCF, SSF, and energy data and shows
comparisons to results in the existing literature.  We offer our
conclusions in Sec.\ \ref{sec:conclusions}.  Hartree atomic units are
used throughout this paper, so that $\hbar= \vert e \vert =m_e=4\pi
\epsilon _0=1$.  

\section{Methods \label{subsec:methods}}

We have used the \textsc{casino}\cite{CASINO2} code to perform variational and
diffusion Monte Carlo\cite{Review} (VMC and DMC) calculations.  In the VMC
method, the Metropolis algorithm is used to generate a set of configurations
distributed according to the square modulus of a trial wave function over
which the local energy is averaged. In the DMC method, an initial wave
function is evolved in imaginary time, which in principle projects out the
ground state.  For fermionic systems, the antisymmetry of the wave function is,
however, imposed via the fixed-node approximation,\cite{FNA} in which the nodal
surface is constrained to remain unchanged during the evolution.  Both the VMC
and DMC methods give an upper bound to the ground-state energy of the system.  
The quality of the parameterization of the trial wave
function and the VMC optimization procedure influence the accuracy of the
results and the statistical noise.

Our wave functions consisted of Slater determinants of plane waves
multiplied by a Jastrow factor and included a backflow
transformation.  The Jastrow factor comprised polynomial and plane-wave
expansions in the electron-electron separation, together with
three-body terms.\cite{Neil-J,Generic-J} The electron coordinates in
the Slater determinant were replaced by quasiparticle coordinates
obtained by a backflow transformation represented by a polynomial in
the electron-electron separation.\cite{Kwon,Inhom-BF}  The variable
parameters in these wave functions were optimized using variance
minimization\cite{Varmin1,Varmin2} and then linear least-squares
energy minimization,\cite{Emin} as this was found to give the most
accurate trial wave functions.

The quality of our wave functions was such that, for unpolarized and
fully polarized systems, the VMC and DMC methods produced essentially
identical PCFs within the statistical precision we were able to
obtain.  For these systems we therefore report DMC expectation values
rather than the extrapolated estimators often used in QMC studies.  In
partially polarized systems, however, our trial wave functions led to
small but statistically significant differences between the VMC and
DMC data for minority-spin electrons.  In these cases we have used
extrapolated estimation,\cite{Extrap} in which the PCF is $g=2g_\text{DMC}-g_\text{VMC}$,
where $g_\text{DMC}$ and $g_\text{VMC}$ are the DMC and VMC PCFs,
respectively. 

We studied 118-electron HEGs in face-centered-cubic simulation cells
and imposed twisted boundary conditions\cite{Lin} so that the
wave function picks up a phase when an electron is translated by a
simulation-cell lattice vector $\mathbf{R}_s$:            
\begin{equation*}
  \Psi (\mathbf{r}_{1},\ldots,\mathbf{r} _i+\mathbf{R}_s,\ldots,\mathbf{r}_N) 
  =  \exp (i \mathbf{k}_s \cdot \mathbf{R}_s) \Psi (\mathbf{r}_1,\ldots,\mathbf{r}_N).
\end{equation*}
The twist offset $\mathbf{k}_s$ was allowed to vary with uniform
probability over the first Brillouin zone of the simulation cell
during our simulations.  Averaging in this way has been shown to 
reduce single-particle finite-size effects in the energy.\cite{TA-energy}

\begin{figure}
\begin{center}
\includegraphics[clip,width=0.45\textwidth]{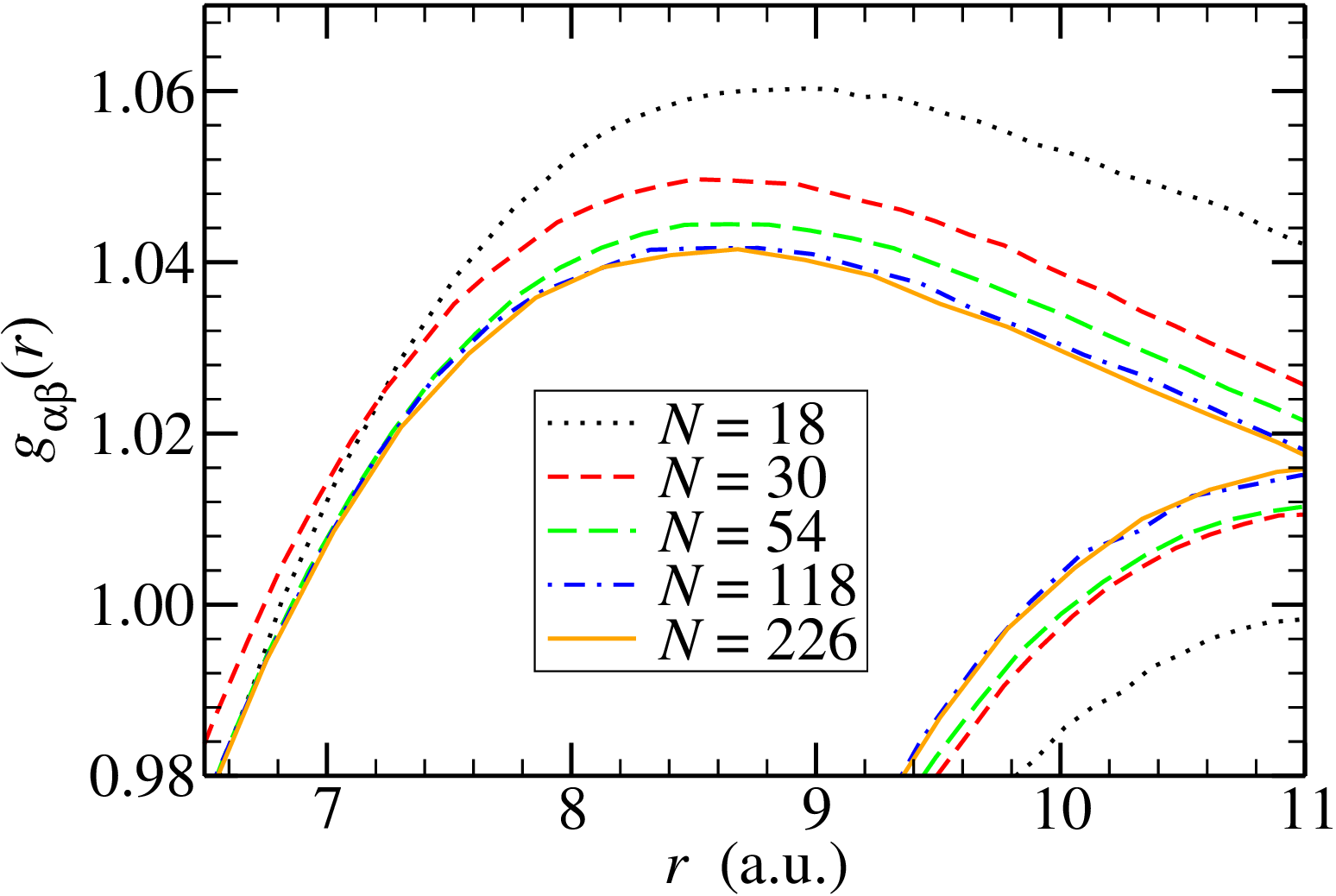}
\caption{(Color online) Finite-size effects in the spin-resolved
  pair correlation function (PCF). The upper curves are for antiparallel 
  spins and the lower-right curves are for parallel spins. The data are 
  from VMC twist-averaged simulations at $r_\text{s}=5$ a.u.\ and $\zeta=0$,
  and at the various sizes shown.  The region around the first peak in
  the antiparallel-spin PCF is shown, where the largest finite-size
  effects occur.  
  \label{fig:finite_size_effects}}
\end{center}
\end{figure}

We investigated the remaining variation with system size in the PCF 
using twist-averaged VMC simulations at $r_\text{s} = 5$ a.u.\ and $\zeta = 0$. 
We show the spin resolved PCF in Fig.\ \ref{fig:finite_size_effects}.  
This figure, like all others in this paper except Fig.\ \ref{fig:fits},
shows our raw QMC data, rather than a fit to the data.  The
region around the first peak in the antiparallel-spin PCF best shows the
variation with system size.  Both $g_{\uparrow \downarrow}(r)$ (upper
curves) and $g_{\uparrow \uparrow}(r)$ (lower-right curves) exhibit
systematic finite-size errors for small systems in this region,
converging to the thermodynamic limit on the scale of the graph at
approximately $N=100$ electrons. (Results
for 118- and 226-electron systems are almost indistinguishable on the scale
of the graph.)  Some cancellation of
errors occurs when the spin-averaged PCF is calculated.

The variation with system size
appears less pronounced and less systematic at smaller $r$. The on-top 
pair density was unaffected within statistical uncertainty.  For systems 
with $N=54$, $118$, and $226$ electrons, there was good agreement at 
small but nonzero distances: typical differences were of order 1--5~\%.  
In contrast, the long-range PCF is affected by unavoidable finite-size 
errors due to the finite size of the simulation cell.  Twist-averaging 
cannot remove the spurious correlation caused by the periodic boundary 
conditions.  This shows up as a small but statistically significant variation 
in the long-range oscillations from about $r/r_s \gtrsim 2$ for 118 electrons 
that decreases in magnitude as the system size increases.
Figure\ \ref{fig:finite_size_effects} is included as an XMGrace file in the 
Supplemental Material\cite{supplemental} accompanying this paper, so that 
other regions of the graph may be inspected.

The energy is known to converge much more slowly with system size than
the PCF\@.\cite{Lin,TA-energy} We have therefore applied analytical
corrections to our energy data following the procedure developed in
Refs.\ \onlinecite{fin_chiesa} and \onlinecite{TA-energy}.  Residual
single-particle finite-size effects are removed by addition of the
difference between the infinite-system and twist-averaged,
finite-system Hartree-Fock kinetic energies.  The leading-order
correction to the Ewald energy, and leading- and next-to-leading-order
corrections to the kinetic energy, are also included.  In addition,
finite-size effects in the energy have previously\cite{Eclr} been
found to be relatively insensitive to $\zeta$, and so a cancellation
of errors renders finite-size effects in the energies of spin
polarization small even without any correction.  The analytical 
arguments of Holzmann \emph{et al.}\cite{HBC} lend further support to this 
conclusion.

We also examined possible time-step and population-control biases and found
them to be very small---less than one part in a thousand of the correlation 
energy in test cases---so we neglect them in what follows. All DMC
calculations were performed using at least 1000 walkers and time steps
for which the acceptance probability was generally greater than
99.7~\%. 

We accumulate the spin-resolved PCF as an average over the set of
configurations generated in the course of the simulation:
\begin{equation*}
  g_{\alpha \beta}(r) = \frac{\Omega}{4\pi r^2 N_{\alpha} N_{\beta}}
  \left< \sum_{j\in\alpha} \sum_{\substack{l\in\beta \\ l\ne j}}
  \delta( \vert \mathbf{r}_j - \mathbf{r}_l \vert - r  ) \right>,
\label{spin-resolved-PCF-defn}
\end{equation*}
where $\Omega$ is the simulation-cell volume and $N_\alpha$ is the number of
electrons of spin $\alpha$.
The SSF is simply related to the Fourier transform of the PCF and is given by 
\begin{equation*}
  S_{\alpha\beta}(\mathbf{k})=\frac{1}{\sqrt{N_{\alpha}N_{\beta}}} \left< \rho_{\alpha}(\mathbf{k}) 
  \rho_{\beta}(\mathbf{-k}) \right> - \sqrt{N_{\alpha}N_{\beta}} \delta_{\mathbf{k0}},
  \label{SSF-defn}
\end{equation*}
where $\rho_{\alpha}(\mathbf{k}) = \sum_{j\in\alpha} \exp
(i\mathbf{k}\cdot\mathbf{r}_j)$. $S_{\alpha\beta}(\mathbf{k})$ is then
spherically averaged in $\mathbf{k}$-space, because the SSF only
depends on the magnitude of $\mathbf{k}$ in a homogeneous and
isotropic system.  

At high densities, the variation in the energy as the twist offset 
is changed can be much greater than other sources of noise in the 
simulation.  Twist-averaging can therefore be computationally expensive 
for high-density HEGs, because a large number of twist angles are 
required in order to obtain a precise energy and the simulation 
must be reequilibrated between twists, preventing rapid changes of twist 
offset in a DMC calculation.  To circumvent this difficulty, we have 
used the Hartree-Fock energy (as a function of twist offset) as a control 
variate.\cite{naval} 

If $E_{\rm QMC}$ denotes the QMC energy, $E_{\rm HFKE}$ 
the Hartree-Fock kinetic energy, and $E_{\rm HFEX}$ the exchange energy as
functions of twist angle, the standard way of getting the twist-averaged energy is 
$\langle E_{\rm QMC} \rangle$, where the angled brackets denote an average over twists.  
Another estimator is 
\begin{equation*}
\theta=E_{\rm QMC} + c_1(E_{\rm HFKE}- \langle E_{\rm HFKE} \rangle ) + c_2(E_{\rm HFEX}- \langle E_{\rm HFEX} \rangle ),
\end{equation*}
where the $\{c_i\}$ are coefficients that can be chosen to minimize the variance of 
$\theta$.  In order to account for the remaining sources of noise, we add the 
correction above to the energy data at each time step and, at the end of the 
simulation, reblock\cite{flyvbjerg} the corrected data to obtain much smaller error bars.

To illustrate the reduction in variance obtained, we apply the method to VMC data 
for two systems in this section: first a HEG with two up-spin electrons and one down-spin electron 
at $r_\text{s}=0.5$ a.u.; then a 118-electron HEG at $r_\text{s}=0.5$ a.u.\ and $\zeta=0$.  
The VMC method allows one to simulate a large number of twist angles in parallel,  
so we are able to compare the corrected data to accurate twist-averages performed 
the standard way. It can be seen in Tables \ref{table:small-system-TA} and 
\ref{table:118-system-TA} that the method successfully reduces the variance 
without introducing bias.  Although we chose twist offsets randomly throughout the 
first Brillouin zone, the method might also improve convergence of twist-averages 
using Monkhorst-Pack\cite{MP} grids.

\begin{table}
\begin{center}
\caption{Energies from VMC simulations in a HEG at $r_\text{s}=0.5$
  a.u.\ containing two up-spin electrons and one down-spin electron.  
The raw VMC energies are compared to the same data after a correction has been
applied using HF kinetic and exchange energies 
as described in the text.  The VMC twist-averaged energy obtained in the standard way, using just 
over $21 \times 10^6$ twist angles, is $3.5484(3)$ a.u.\ per electron.
\label{table:small-system-TA}}
\begin{tabular}{lr@{.}lc}
\hline \hline

& \multicolumn{3}{c}{VMC energy (a.u./elec.)} \\

\raisebox{1.5ex}[0pt]{\# twists} & \multicolumn{2}{c}{Raw} & Corr. \\

\hline

$10$           & $3$&$73(40)$ & $3.5480(1)$ \\

$20$           & $3$&$57(39)$ & $3.5480(1)$ \\

$100$          & $3$&$42(10)$ & $3.5480(1)$ \\

$441$          & $3$&$51(6)$  & $3.5480(1)$ \\

\hline \hline
\end{tabular}
\end{center}
\end{table}

\begin{table}
\begin{center}
\caption{Energies from VMC simulations in a 118-electron HEG at $r_\text{s}=0.5$ a.u.\ and $\zeta=0$.
The raw VMC energies are compared to the same data after a correction has been
applied using HF kinetic and exchange energies
as described in the text.  The VMC twist-averaged energy obtained in the standard way, using 
$4.8 \times 10^6$ twist angles, is $3.41378(2)$ a.u.\ per electron.
\label{table:118-system-TA}}
\begin{tabular}{lcc}
\hline \hline

& \multicolumn{2}{c}{VMC energy (a.u./elec.)} \\

\raisebox{1.5ex}[0pt]{\# twists} & Raw & Corr. \\

\hline

$10$           & $3.424(4)$  & $3.41377(4)$ \\

$20$           & $3.424(4)$  & $3.41375(4)$ \\

$50$           & $3.414(5)$  & $3.41380(3)$ \\

$156$          & $3.416(2)$  & $3.41377(2)$ \\

\hline \hline
\end{tabular}
\end{center}
\end{table}

\section{Results \label{sec:results}}

\subsection{Pair Correlation Function \label{subsec:PCF}}

\begin{figure}
\begin{center}
\includegraphics[clip,width=0.45\textwidth]{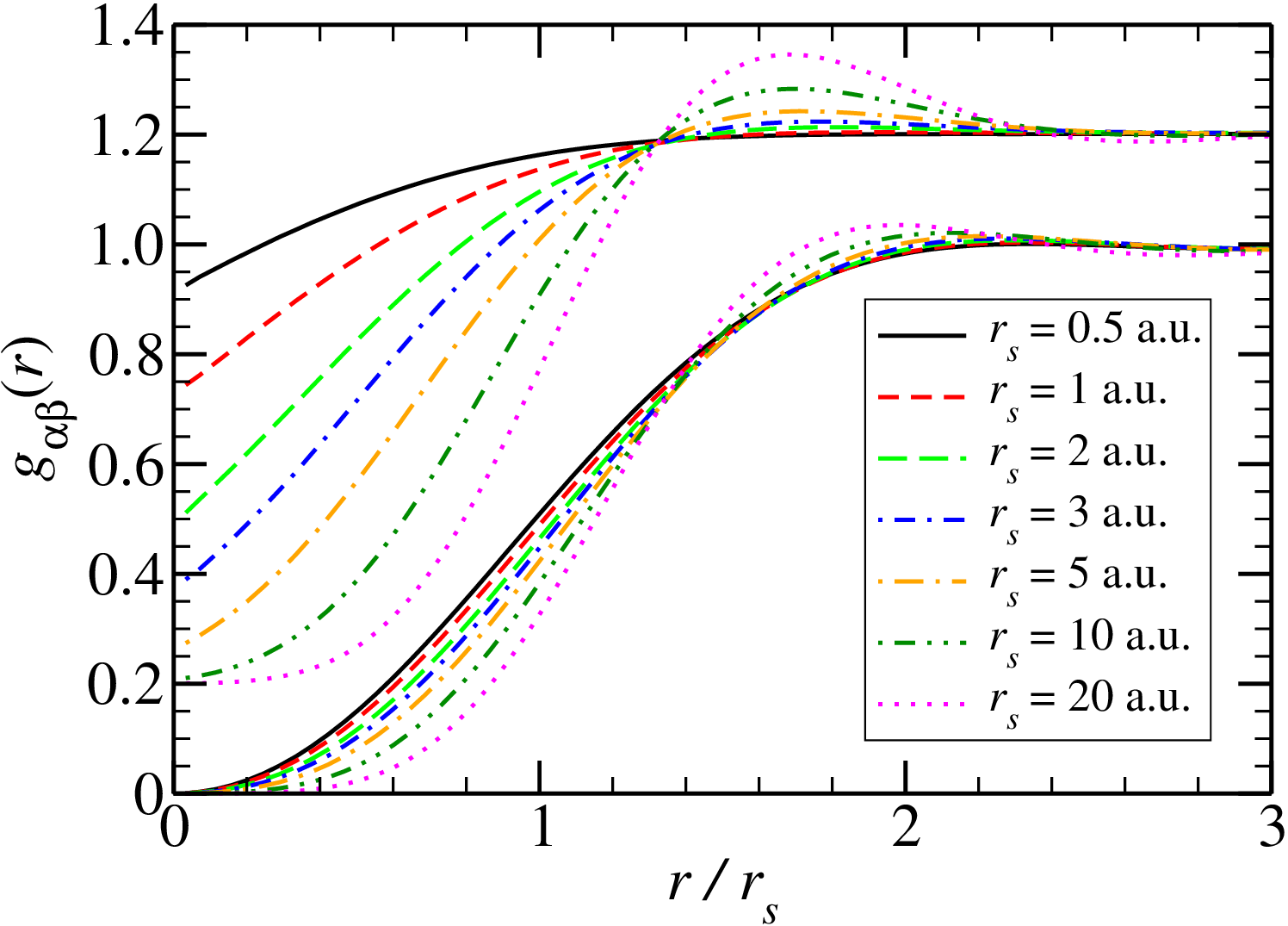}
\caption{(Color online) Spin-resolved pair correlation functions (PCFs) for 
  unpolarized HEGs calculated at the densities shown.  The antiparallel-spin 
  PCFs are translated upwards by 0.2 units.  The data were obtained using 
  twist-averaged DMC simulations.
\label{fig:PCFs-vs-rs}}
\end{center}
\end{figure}

Spin-resolved pair correlation functions (PCFs) are shown in 
Fig.\ \ref{fig:PCFs-vs-rs} for the unpolarized HEG at different densities 
using twist averaging.  Antiparallel-spin PCFs are translated upwards by 0.2 
units for greater visibility, and show much greater variation with density.  
For $\zeta=0$ (and $\zeta=1$, not shown) the VMC and DMC results are very
similar, so in the plot we present only DMC data.

\begin{figure}
\begin{center}
\includegraphics[clip,width=0.45\textwidth]{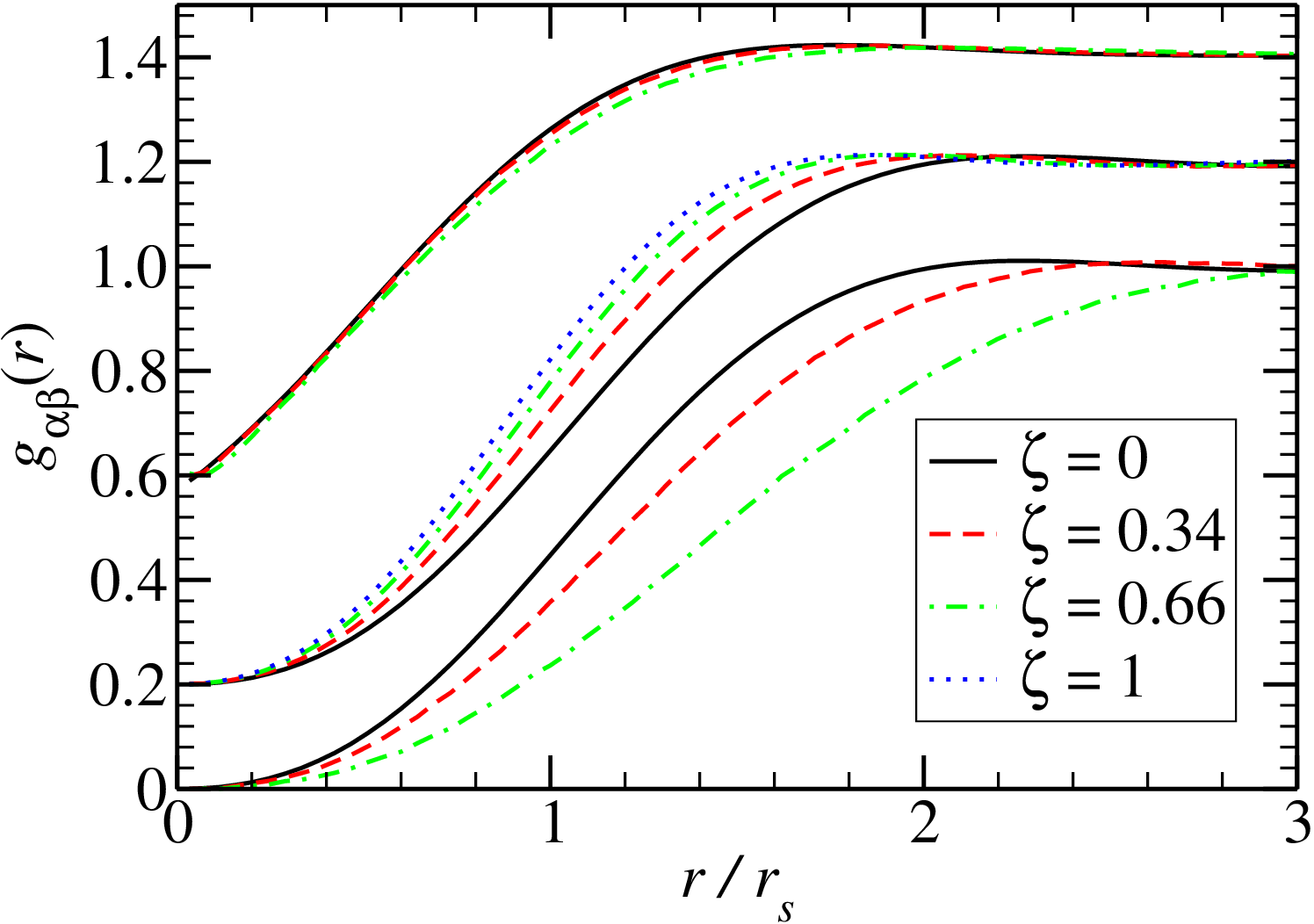}
\caption{(Color online) Variation in the pair correlation function (PCF) 
with spin polarization at $r_\text{s}=3$ a.u.  Antiparallel-spin PCFs 
are translated upwards by 0.4 units, parallel-majority-spin PCFs are 
translated upwards by 0.2 units, and the lower curves give the 
parallel-minority-spin PCFs.  Note that the $\zeta=0$ parallel-spin PCF 
is shown twice (in black) for comparison with both up-up and down-down 
spin PCFs in those cases where they differ.  The DMC data are shown for 
unpolarized and fully polarized systems, while extrapolated estimates are 
used for intermediate polarizations.  All of the data are twist-averaged.
\label{fig:PCFs-vs-zeta}}
\end{center}
\end{figure}

We show the variation of the PCFs with spin polarization at
$r_\text{s}=3$ a.u.\ in Fig.\ \ref{fig:PCFs-vs-zeta}.  In this figure,
we translate the antiparallel-spin PCFs upwards by 0.4 units and the
parallel-majority-spin PCFs upwards by 0.2 units.  For $\zeta=0$ and
$\zeta=1$, we have used the DMC values. For $\zeta=0.34$ and
$\zeta=0.66$, the extrapolated estimate is shown, because small but
statistically significant differences emerged between the DMC and VMC
data in the parallel-minority-spin PCFs.  
Compared with the parallel-spin PCFs, the antiparallel-spin PCF is
relatively insensitive to the degree of spin polarization.

\begin{figure}
\begin{center}
\includegraphics[clip,width=0.45\textwidth]{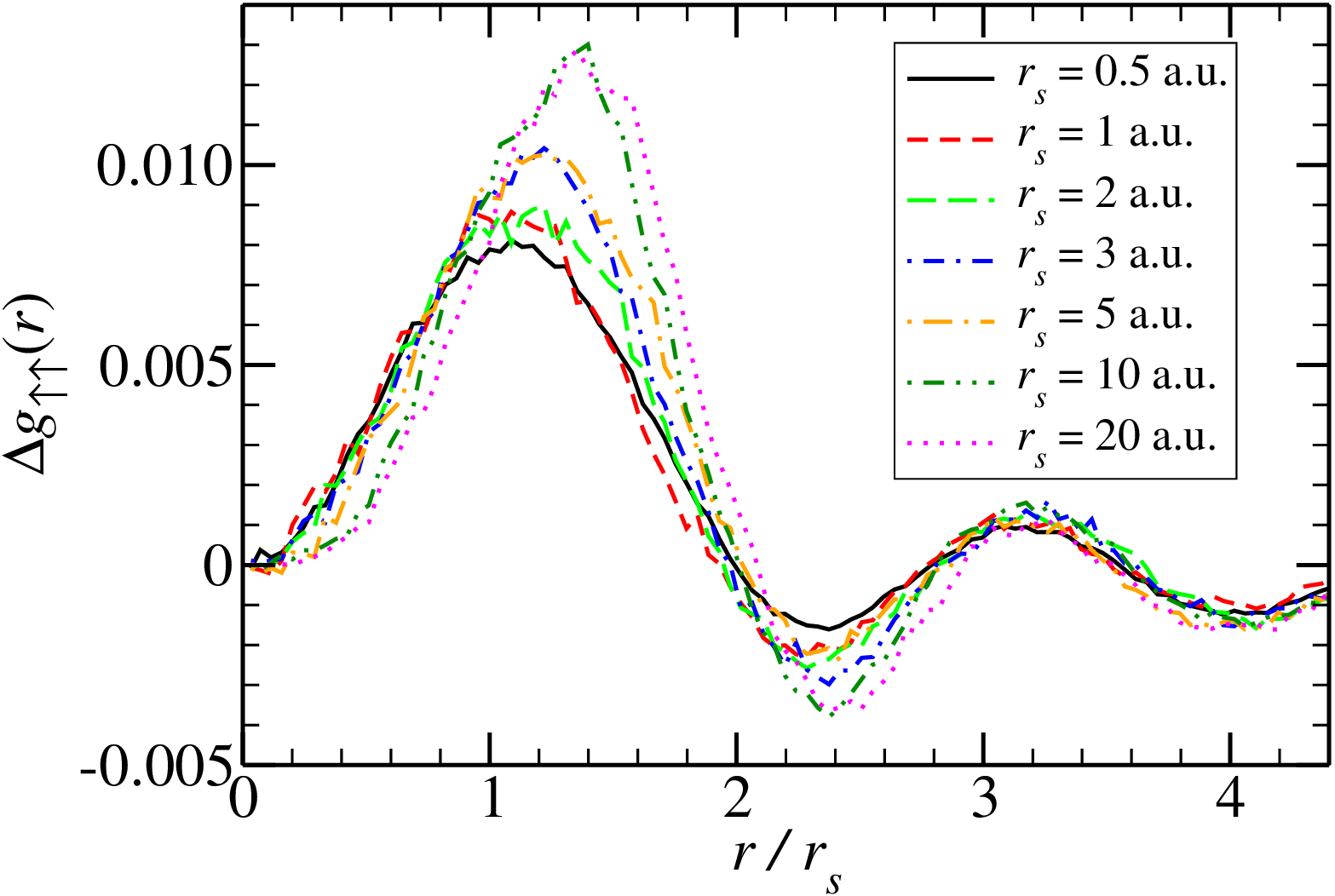}
\caption{(Color online) Changes in the pair correlation function (PCF) due 
to twist averaging.  We plot parallel-spin PCFs with twist averaging minus 
the same quantity evaluated without twist averaging, for the densities shown 
and $\zeta=0$.  Only DMC results are shown, as the VMC results are very similar.
\label{fig:TA-minus-NOTA-PCF-parallel}}
\end{center}
\end{figure}

The PCF data discussed above are twist-averaged to reduce single-particle
finite-size effects.  Twist averaging has a significant
effect on the energy data, but the effect on the PCF data is much
smaller.  We show in Fig.\ \ref{fig:TA-minus-NOTA-PCF-parallel} the
difference $g_\text{TA}-g_\text{PBC}$ between imposing twist-averaged and
periodic boundary conditions for parallel spins.  Noise
obscures any clear pattern for the antiparallel spins, although there
is a tendency for the equivalent figure to show a small dip before
rising to zero, so that $g(0)$ appears to be largely unaffected.

\begin{figure}
\begin{center}
\includegraphics[clip,width=0.45\textwidth]{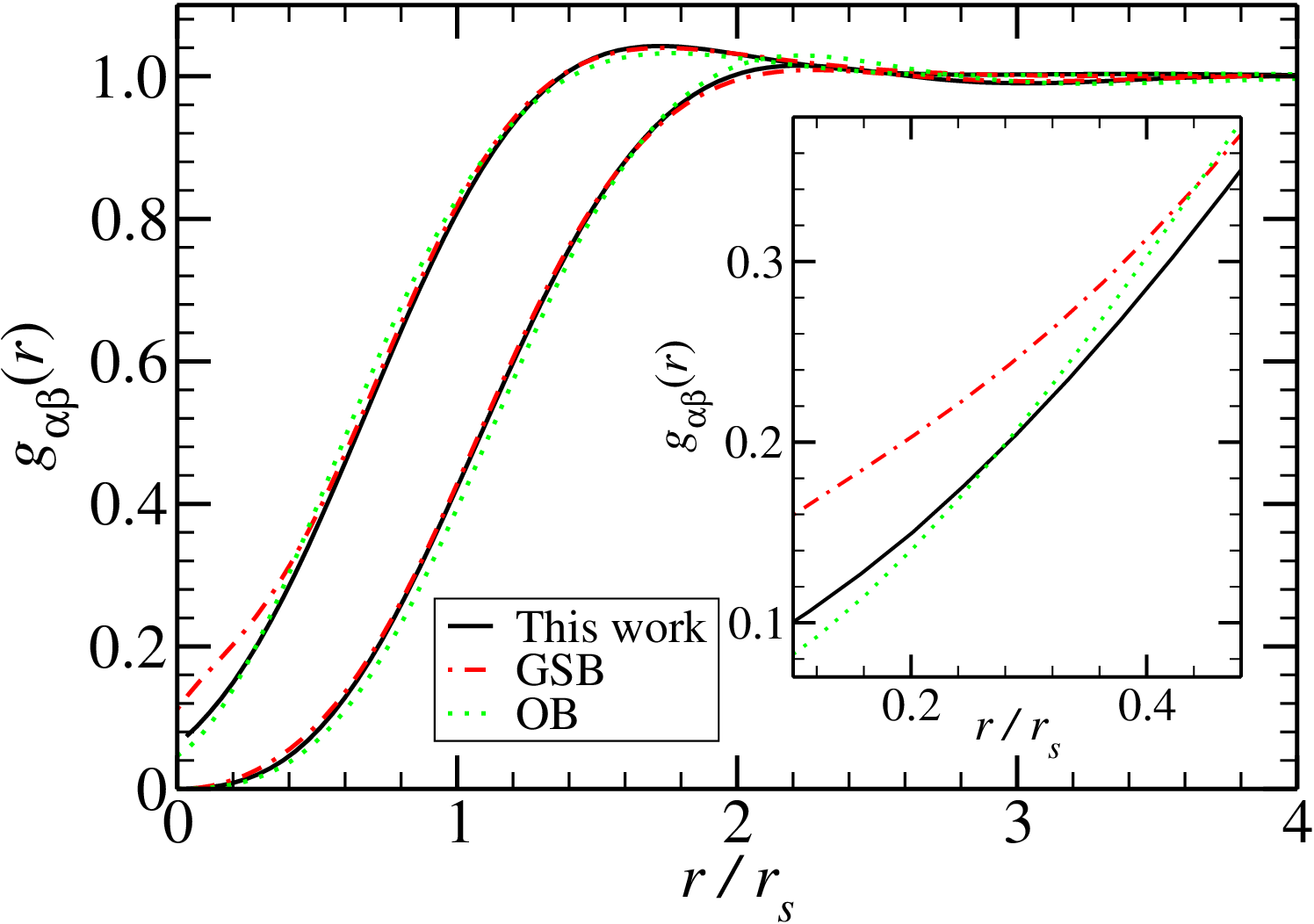}
\caption{(Color online) Spin-resolved pair correlation functions (PCFs) for 
  unpolarized HEGs at $r_\text{s}=5$ a.u.  We show our raw twist-averaged 
  DMC data, together with the fit to DMC results from Ortiz and Ballone\cite{OB94} 
  (labeled OB) given in their paper, and a fit due to Gori-Giorgi,
  Sacchetti, and Bachelet\cite{GSB} using the DMC data of Ortiz, Harris,
  and Ballone,\cite{OHB} (labeled GSB)\@.  The inset shows the low-$r$
  region, where the differences are largest.
  \label{fig:OB-vs-GSB-vs-GGS}}
\end{center}
\end{figure}

We compare our results at $r_\text{s}=5$ a.u.\ and $\zeta=0$ with QMC data
from the literature in Fig.\ \ref{fig:OB-vs-GSB-vs-GGS}.  We find good
agreement with the data of Gori-Giorgi \emph{et al.}\cite{GSB}\ for
intermediate $r/r_\text{s}$.  Our on-top pair densities are closer to
those of Ortiz and Ballone\cite{OB94} as shown in the inset, although
at intermediate $r/r_\text{s}$ modest (of order 5~\%) differences are
obtained.  The figure shows our raw QMC data, whereas the other curves are 
fits to the QMC data of Ortiz, Harris, and Ballone,\cite{OHB} and 
Ortiz and Ballone\cite{OB94} respectively.  Both of these fits obey the 
Kimball cusp conditions and we are confident our QMC data are also 
consistent with these conditions.  As mentioned below, our short-range PCFs 
are well-described by a quadratic of the form 
$g_{\uparrow \downarrow}(r)=a+ar+br^2$, which satisfies the first Kimball condition.

\begin{figure}
\begin{center}
\includegraphics[clip,width=0.45\textwidth]{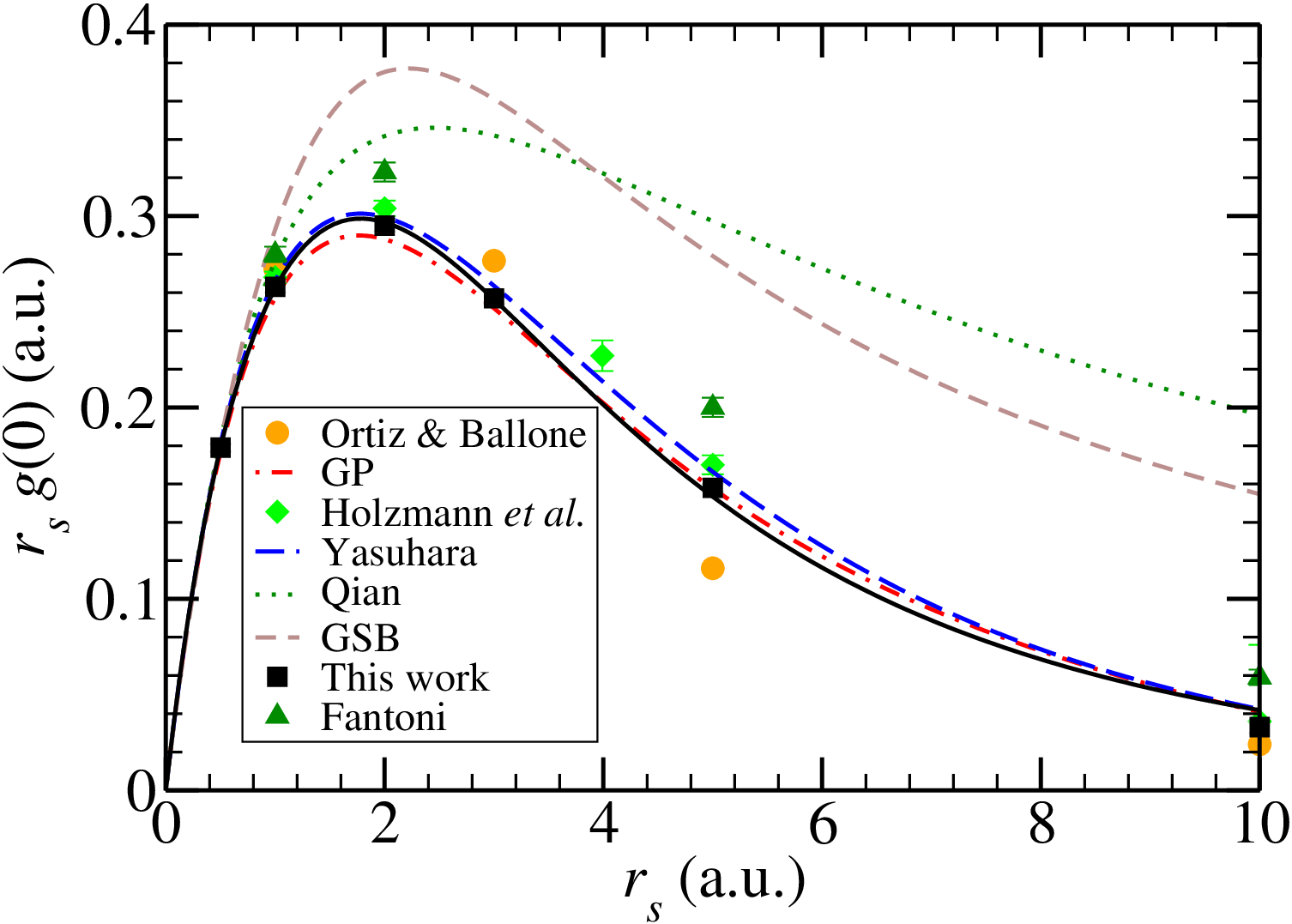}
\caption{(Color online) The on-top pair density $g(0)$ (multiplied by
  $r_\text{s}$) as a function of $r_\text{s}$ for unpolarized systems.
  The data shown are those of: our averaged DMC and VMC results, together 
  with the fit of Eq.\ (\ref{eq:fit-on-top}) (solid line);
  Gori-Giorgi and Perdew;\cite{GP2} Holzmann \emph{et
    al.};\cite{Holzmann} Yasuhara;\cite{Yasuhara} Qian;\cite{Qian}
  Gori-Giorgi \emph{et al}.; \cite{GSB} Ortiz and Ballone;\cite{OB94} 
  and Fantoni.\cite{Fantoni} Error bars are shown for our
  data, those of Holzmann \emph{et al.},\cite{Holzmann} and 
  Fantoni,\cite{Fantoni} but they
  are mostly smaller than the size of the symbols.  Our values are
  slightly smaller than those of Holzmann \emph{et al.}\cite{Holzmann}
  \label{fig:g(0)-unpolarised}}
\end{center}
\end{figure}

We show in Fig.\ \ref{fig:g(0)-unpolarised} our $g(0)$ values for
the unpolarized system in comparison to those in the literature. We have 
averaged our DMC and VMC data in this figure to reduce statistical 
uncertainty, because the $\zeta=0$ PCFs produced by the two methods are identical 
within the statistical precision we were able to obtain.  We used the 
above quadratic form to estimate $g(0)$ from our PCF data at 
finite distances, $g(r)$, and obtained good fits for all systems.  Our
results are in good agreement with the recent QMC data of Holzmann
\emph{et al.},\cite{Holzmann} and the results of Gori-Giorgi and
Perdew\cite{GP2} who used various data including QMC results, and the results of
Yasuhara.\cite{Yasuhara} Our calculations give significantly smaller
values of $g(0)$ than those of Gori-Giorgi, Sacchetti, and
Bachelet,\cite{GSB} and those of Qian.\cite{Qian} This may suggest
that Yasuhara's scheme successfully incorporates screening effects
into the ladder theory calculations.
Nagy \emph{et al.}\cite{Nagy}\ suggest that the description of 
screening used by Gori-Giorgi and Perdew is also an important factor 
in explaining the success of the latter authors' model.

\begin{figure}
\begin{center}
\includegraphics[clip,width=0.45\textwidth]{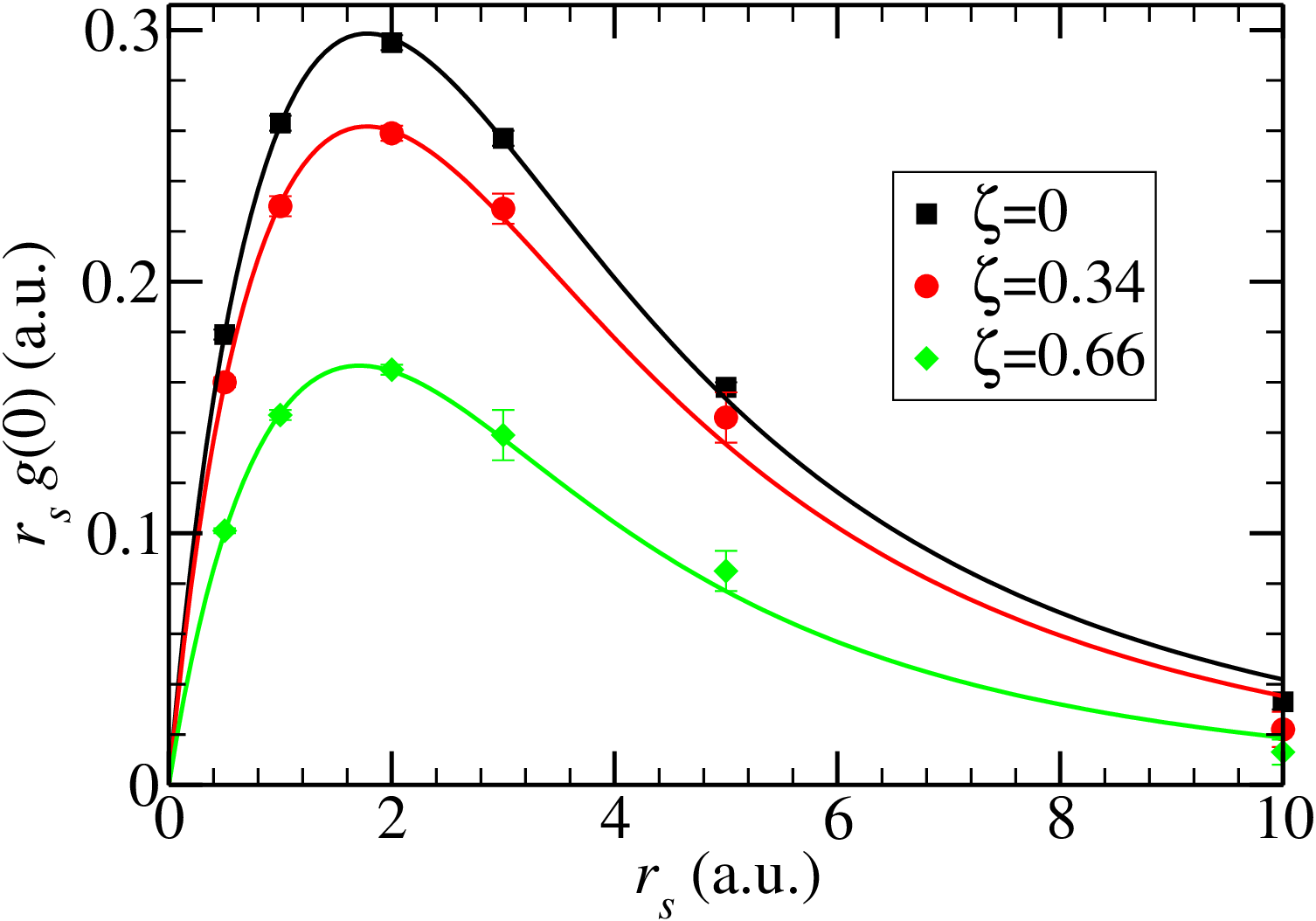}
\caption{(Color online) The on-top pair density (multiplied by
  $r_\text{s}$) as a function of $r_\text{s}$ and $\zeta$.  Averaged DMC
  and VMC data are used for the unpolarized systems, and extrapolated 
  estimates for the polarized systems.  Twist averaging was used for all 
  data shown.  An error bar is plotted for each data point, although some 
  of them are smaller than the size of the symbols. The partially polarized
  systems show a larger statistical uncertainty.  The lines are fits
  as discussed in the text.
  \label{fig:g(0)-polarised}}
\end{center}
\end{figure}

We have also obtained on-top pair densities for partially polarized
systems; see Fig.\ \ref{fig:g(0)-polarised}.  In this case we used
extrapolated estimation due to the small but statistically significant
differences between our VMC and DMC results, although these
differences are much less pronounced in the antiparallel-spin PCF
than in the parallel-minority-spin PCF\@.  This increases the
statistical uncertainty in our results, an effect compounded by the
increasing sparsity of sampling antiparallel spin coalescences as the polarization
increases.  
Nevertheless, as expected from Fig.\
\ref{fig:PCFs-vs-zeta}, the variation of $g(0)$ with $\zeta$ arises
largely from changes in the weights in Eq.\ (\ref{eq:weighted-average})
rather than in the spin-resolved PCFs themselves.  We fit our $g(0)$
data to the following parameterized form in the density range $0.5 \leq
r_\text{s} \leq 20$ a.u.:
\begin{equation}
\label{eq:fit-on-top}
g_{\uparrow \downarrow}(0;r_\text{s})=\frac{1+a\sqrt{r_\text{s}}+br_\text{s}}{1+cr_\text{s}+dr_\text{s}^3},
\end{equation}
and we list the optimal fit parameters obtained in Table\
\ref{table:on-top-parameters}.  This functional form obeys the exact 
high density limit and fits our data well up to $r_\text{s}=20$ a.u., 
although at $r_\text{s}=10$ a.u., it tends to give slightly higher 
values than our QMC calculations obtained, typically by about 2 standard 
deviations.  

\begin{table}
\begin{center}
\caption{
Parameters (in a.u.)\ obtained from fits to the on-top pair density as a function of $r_\text{s}$
using Eq.\ (\ref{eq:fit-on-top}).  The data for the unpolarized systems are averaged DMC 
and VMC results; extrapolated estimation is used for the others.  
\label{table:on-top-parameters}}
\begin{tabular}{lr@{.}lr@{.}lr@{.}lc}
\hline \hline

$\zeta$ & \multicolumn{2}{c}{$a$} & \multicolumn{2}{c}{$b$} & \multicolumn{2}{c}{$c$} & $d$ \\

\hline

$0$           & $0$&$18315$ & $-0$&$0784043$ & $1$&$02232$ & $0.0837741$ \\

$0.34$        & $0$&$284118$ & $-0$&$110062$ & $1$&$1618$  & $0.0874753$ \\

$0.66$        & $0$&$0659538$ & $-0$&$0590569$ & $0$&$836458$ & $0.0832258$ \\

\hline \hline
\end{tabular}
\end{center}
\end{table}

\subsection{Static Structure Factor \label{subsec:SSF}}

\begin{figure}
\begin{center}
\includegraphics[clip,width=0.45\textwidth]{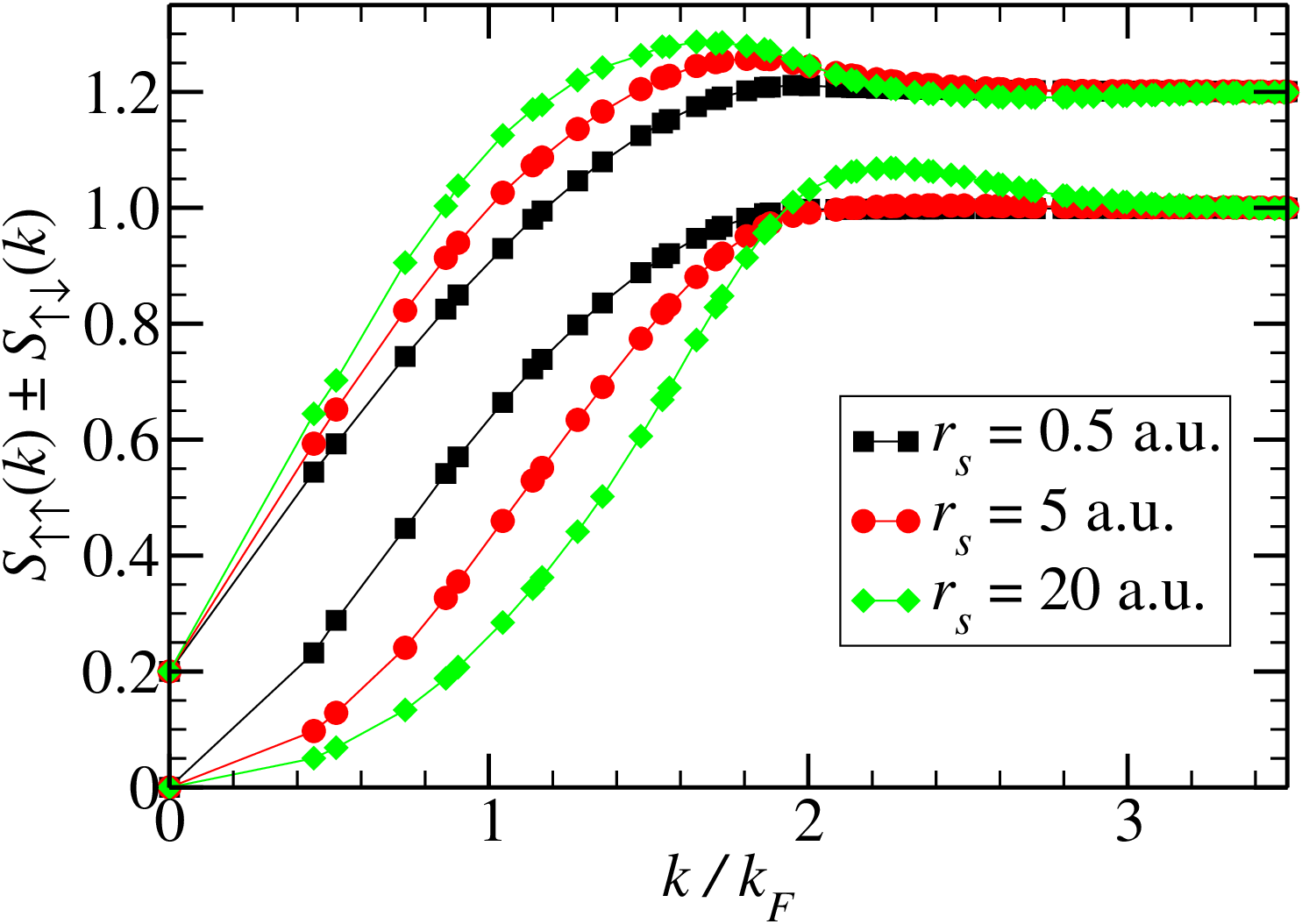}
\caption{(Color online) Structure factors at $\zeta=0$ for three densities.  
The upper curves show $S_{\uparrow \uparrow} - S_{\uparrow \downarrow}$ and 
are translated upwards by 0.2 units; the lower curves show 
$S_{\uparrow \uparrow} + S_{\uparrow \downarrow}$.  The lines are visual guides 
only. The statistical uncertainty is smaller than the size of the symbols.  The 
DMC method with twist averaging was used, as the VMC data are again very similar.  
Intermediate densities lie between the curves shown.
\label{fig:SSF-vs-rs}}
\end{center}
\end{figure}

Static structure factors (SSFs) with $\zeta=0$ at three representative 
densities are shown in
Fig.\ \ref{fig:SSF-vs-rs}.  The upper curves show $S_{\uparrow
  \uparrow} - S_{\uparrow \downarrow}$ and the lower curves,
$S_{\uparrow \uparrow} + S_{\uparrow \downarrow}$.  The differences
between the DMC and VMC results are again smaller than the statistical
noise, and so we plot DMC values.  The parallel-spin structure factors 
change very slowly with density, and the $r_\text{s}$-dependence shown arises
almost entirely from the antiparallel-spin structure factor, which is 
(mostly) negative and of increasing magnitude as the density decreases.

\begin{figure}
\begin{center}
\includegraphics[clip,width=0.45\textwidth]{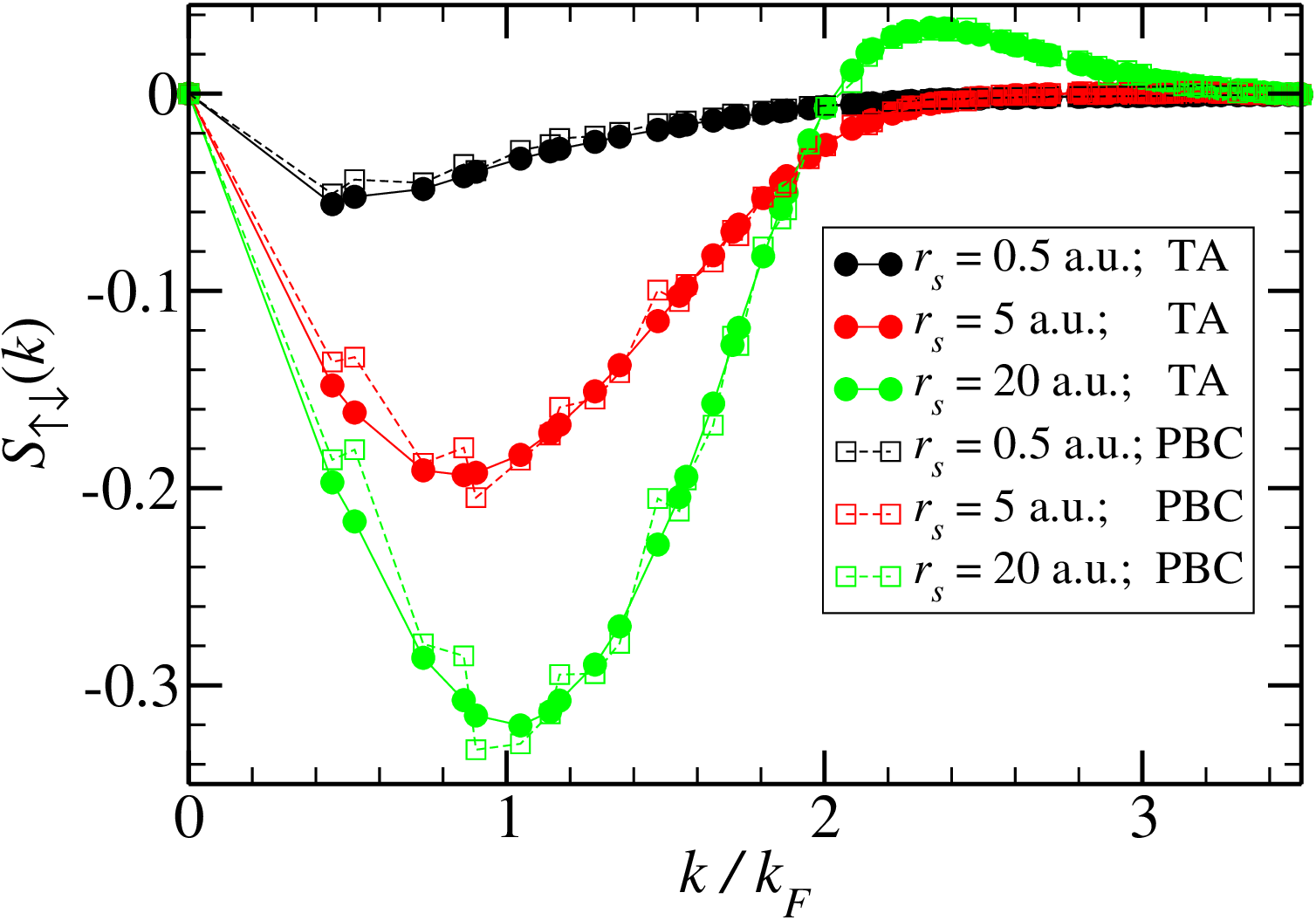}
\caption{(Color online) Effect of twist averaging on the
  antiparallel-spin static structre factor (SSF).  We plot SSFs at $\zeta=0$ 
  for three densities, with twist averaging (TA) and without (PBC)\@.  Higher
  densities are towards the top of the graph; intermediate densities
  lie between the data shown and similar effects are seen in
  parallel-spin SSFs.  Only DMC data are plotted as the VMC data are
  very similar.
\label{fig:TA-vs-NOTA-SSF}}
\end{center}
\end{figure}

Twist averaging was used to obtain the results shown in
Fig.\ \ref{fig:SSF-vs-rs}, and its effect was
more pronounced than in the PCFs, as shown in Fig.\
\ref{fig:TA-vs-NOTA-SSF}, where we plot antiparallel-spin SSFs at $\zeta=0$
with and without twist averaging for the same three densities.  Shell-filling
effects are clearly visible in the non-twist-averaged data.
The VMC and DMC results are very similar and so we can be confident
that the differences do not arise from statistical noise.  Small finite-size 
effects are visible in the low-$ \vert \mathbf{k} \vert $ structure factor 
data.  These correspond to the inevitable finite-size effects in the 
long-range part of the PCF mentioned above.

\subsection{Energies \label{subsec:Energies}}

Twist-averaged DMC energies are reported in Table\
\ref{table:Energy-vs-rs-zeta} for $r_\text{s}=0.5$--20 a.u.\ and for
$\zeta=0$, 0.34, 0.66, and 1.  The data are for finite (small)
time steps such that the DMC acceptance probability was almost always
greater than 99.7\%. We verified in sample cases that the
time-step and population-control biases were small: typically less 
than one part in a thousand of the correlation energy.  The finite-size 
effects in the energy are substantial for the 118-electron HEGs, so we 
have corrected for these, as discussed in Sec.\ \ref{subsec:methods}.  
The data are further corrected using the 
control variate method, also discussed in Sec.\ \ref{subsec:methods}, 
because twist-averaging would otherwise lead to large error bars, 
especially at high densities.

We find that the ground state is unpolarized for all densities studied
here, in agreement with the most recent calculations.\cite{Lin} This
is also clearly evident in our VMC results, despite the VMC method
suffering a bias towards polarized systems due to the relative ease with which
correlation effects in the wave function may be parameterized.  This bias
decreases with increased variational freedom in the trial wave function.

\begin{figure}
\begin{center}
\includegraphics[clip,width=0.45\textwidth]{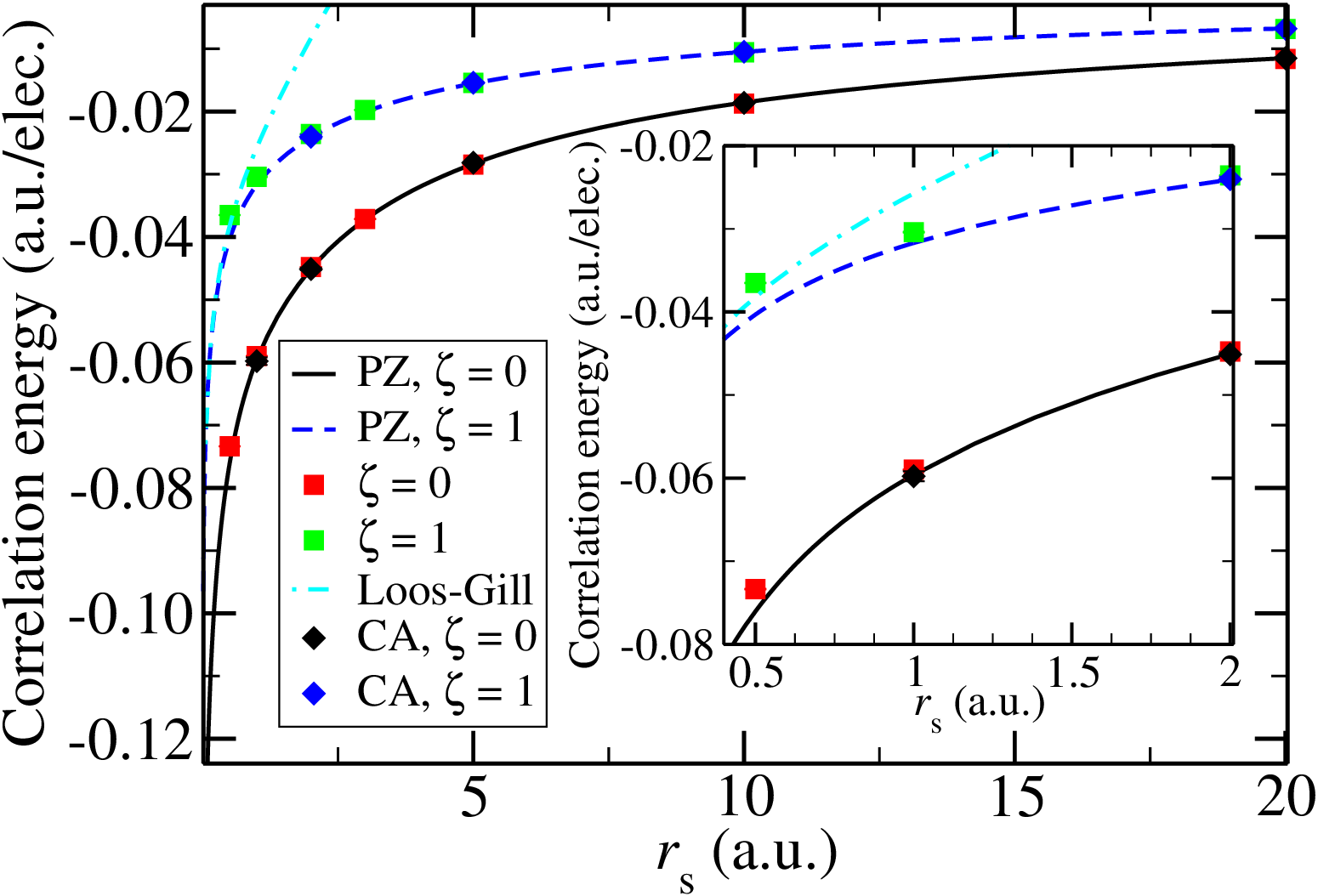}
\caption{(Color online) 
Correlation energy at $\zeta=0$ and 1 against density parameter $r_\text{s}$.  
The present work is simply labeled ``$\zeta=0$'' and ``$\zeta=1$''.  Also 
shown are the Ceperley-Alder QMC data;\cite{CepAlder} the Perdew-Zunger fit 
to the latter data;\cite{PZ} and the recent high-density result of 
Loos and Gill.\cite{LG}  Our data are corrected for finite-size effects.  
Time step extrapolation was not performed.  The statistical error bars are 
shown, but are usually smaller than the symbols.
\label{fig:Loos-Gill-Energy-2}}
\end{center}
\end{figure}

Our energies for unpolarized and fully polarized HEGs are shown in 
Fig.\ \ref{fig:Loos-Gill-Energy-2}, where we also plot for comparison 
the Ceperley-Alder data\cite{CepAlder} and the fit to the latter obtained by Perdew 
and Zunger,\cite{PZ} as used in the LSDA\@.  In addition, a recent high-density 
(RPA) limit for the polarized system due to Loos and Gill\cite{LG} is shown.  
The present results follow the Perdew-Zunger fit closely, except at 
high density, where we find slightly smaller correlation energies, 
especially at full polarization.  The finite-size corrections we 
have applied are significantly larger than the differences shown in 
Fig.\ \ref{fig:Loos-Gill-Energy-2}, and it is possible that higher 
order finite-size corrections would account for the difference in 
energies obtained.  It is worth noting that Ceperley and Alder did 
not perform QMC calculations at densities higher than $r_\text{s}=2$ a.u.\ 
for $\zeta=1$ and that the Loos-Gill result also gives smaller 
correlation energies in this region than the Perdew-Zunger fit.

To compare our intermediate-$\zeta$ energies with those in the 
literature, we used the procedure developed by Perdew and 
Zunger to interpolate between $\zeta=0$ and $\zeta=1$.  We used 
our QMC data for the unpolarized and fully polarized systems and 
interpolated using Eq.\ (C12) in their paper.  These fits are included as 
part of the Supplemental Material accompanying this paper.\cite{supplemental}  The interpolation scheme is 
very successful at low density, whereas our high-density data for 
intermediate polarizations are not reproduced with such high accuracy. 

\begin{figure}
\begin{center}
\includegraphics[clip,width=0.45\textwidth]{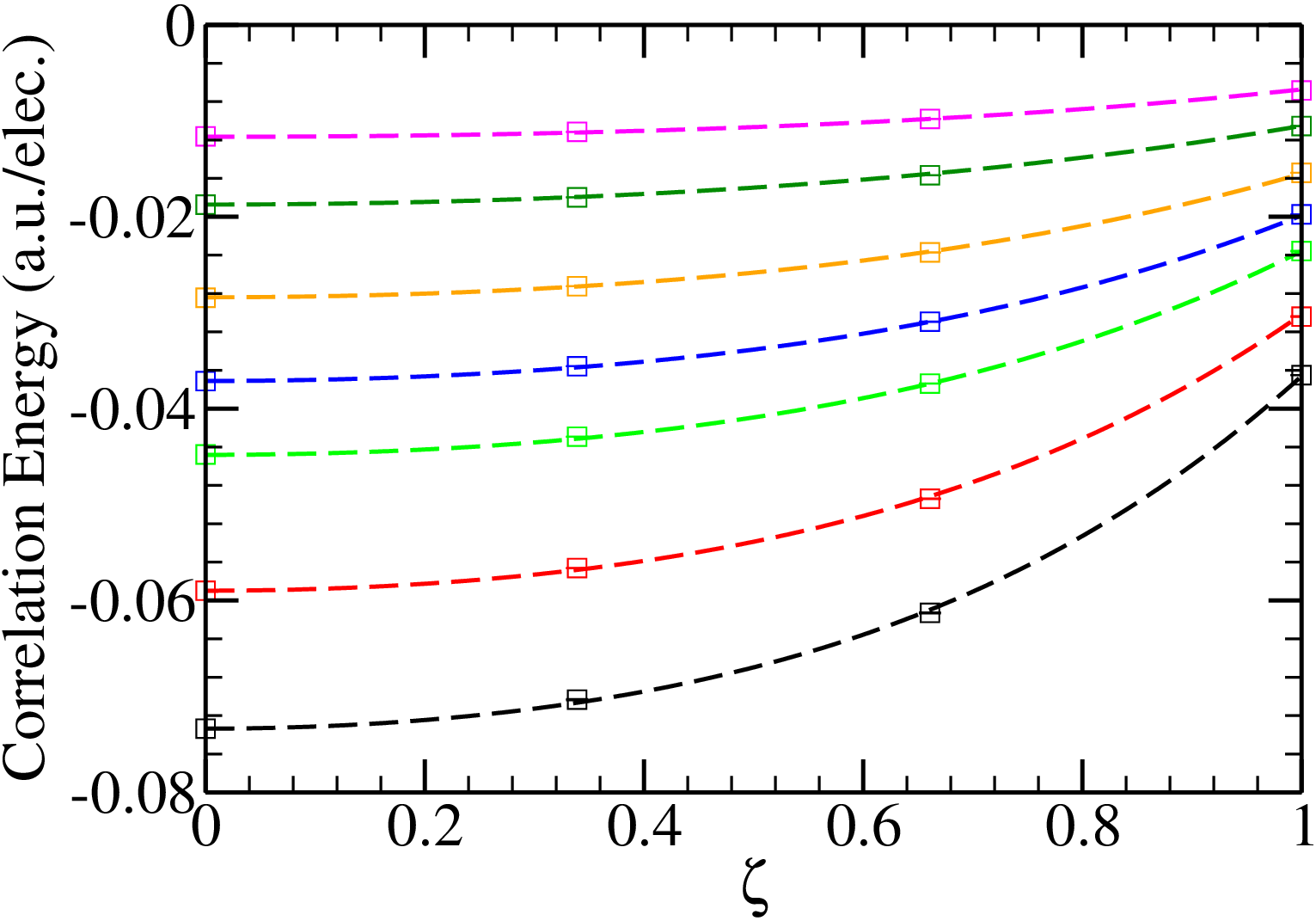}
\caption{(Color online) Correlation energies obtained in DMC calculations together with the fit of 
Eqs.\ (\ref{eq:functional}), (\ref{eq:f_i}), and (\ref{eq:Xi}).  Dashed lines show the fit; square symbols the QMC data.  Lower densities 
appear towards the top of the graph; higher densities towards the bottom of the figure.  From 
top-to-bottom, therefore, the densities shown are: $r_\text{s}=20$, 10, 5, 3, 2, 1, and 0.5 a.u.\  
Error bars on the QMC data are shown, but are smaller than the size of the symbols.
Corrections have been applied for finite-size effects.  Time-step extrapolation was not performed.
\label{fig:PZ-LSDA-fit3}}
\end{center}
\end{figure}

In Fig.\ \ref{fig:PZ-LSDA-fit3} we show our correlation energies 
together with a quartic fit in $\zeta$ at each density:
\begin{multline}
  E_c(r_\text{s},\zeta) = f_0(r_\text{s}) + \Xi(r_\text{s}) \Delta
  f(r_\text{s}) \zeta^2 \\ + \left[ 1-\Xi(r_\text{s}) \right] \Delta f(r_\text{s}) \zeta^4,
\label{eq:functional}
\end{multline}
where $\Delta f(r_\text{s})=f_1(r_\text{s})-f_0(r_\text{s})$ and  $f_\zeta(r_\text{s})$ denotes 
the correlation energy at spin polarization $\zeta$.  We fit
$f_\zeta(r_\text{s})$ for $\zeta=0$ and $1$ over the density range
$r_\text{s}=0.5$--$20$ a.u.\ using the functional form 
\begin{equation}
f_i(r_\text{s})=\frac{\gamma_i}{1+\beta_1^ir_\text{s}^\frac{1}{2}+\beta_2^ir_\text{s}}
\label{eq:f_i}
\end{equation}
suggested by Ceperley,\cite{Cep} and Perdew and Zunger.\cite{PZ}  We find the interpolation
\begin{equation}
\Xi(r_\text{s})=a+br_\text{s}+cr_\text{s}^2
\label{eq:Xi}
\end{equation}
gives a good fit to the intermediate-$\zeta$ QMC data.  The energies can be 
extrapolated to the high-density limit using the method of Perdew and Zunger,\cite{PZ} by matching the correlation energy and potential arising from $f_i$ 
to the high-density expansions at $r_\text{s}=0.5$ a.u., although it is not clear 
if the interpolation between unpolarized and polarized systems used 
here will be reliable at higher densities. 

All data shown in Fig.\ \ref{fig:PZ-LSDA-fit3} have been corrected for
finite-size effects, as discussed above.  A control variate correction has 
also been applied to reduce the
statistical uncertainty arising from the finite sample of twist
angles, as discussed in Sec.\ \ref{subsec:methods}.  The optimum
parameter values used in the fit of Eqs.\ (\ref{eq:functional}),
(\ref{eq:f_i}), and (\ref{eq:Xi}) are given in Table
\ref{table:functional-parameters}. 

\begingroup
\squeezetable
\begin{table}
\begin{center}
\caption{Energies as a function of density parameter and spin polarization.  The 
DMC data are twist-averaged and include corrections for finite-size effects, as 
discussed in the text.  Time-step extrapolation was not performed.  Data include 
a control variate correction.
\label{table:Energy-vs-rs-zeta}}
\begin{tabular}{lr@{.}lr@{.}lr@{.}lr@{.}l}
\hline \hline

& \multicolumn{8}{c}{Energy (a.u./elec.)} \\

\raisebox{1.5ex}[0pt]{$r_\text{s}$ (a.u.)} & \multicolumn{2}{c}{$\zeta=0$} & \multicolumn{2}{c}{$\zeta=0.34$}  & \multicolumn{2}{c}{$\zeta=0.66$} & \multicolumn{2}{c}{$\zeta=1$} \\

\hline

$0.5$   &  $3$&$43011(4)$      &  $3$&$69287(6)$      &  $4$&$44164(6)$      &  $5$&$82498(2)$ \\

$1$     &  $0$&$58780(1)$      &  $0$&$64919(2)$      &  $0$&$82394(4)$      &  $1$&$14634(2)$ \\
 
$2$     &  $0$&$002380(5)$     &  $0$&$016027(6)$     &  $0$&$05475(2)$     &  $0$&$12629(3)$  \\

$3$     &  $-0$&$067075(4)$    &  $-0$&$061604(5)$    &  $-0$&$04608(2)$    &  $-0$&$017278(4)$ \\

$5$     &  $-0$&$075881(1)$   &  $-0$&$074208(4)$    &  $-0$&$069548(4)$   &  $-0$&$060717(5)$ \\

$10$    &  $-0$&$0535116(5)$   &  $-0$&$053214(2)$   &  $-0$&$052375(2)$   &  $-0$&$0507337(5)$ \\

$20$    &  $-0$&$0317686(5)$  &  $-0$&$0317156(7)$  &  $-0$&$0315940(7)$  &  $-0$&$0313160(4)$ \\

\hline \hline
\end{tabular}
\end{center}
\end{table}
\endgroup

\begin{table}
\begin{center}
\caption{
Parameters (in a.u.)\ obtained from the fit to the correlation energy as a function of $r_\text{s}$
and $\zeta$ using Eq.\ (\ref{eq:functional}).  
\label{table:functional-parameters}}
\begin{tabular}{lr@{.}l}
\hline \hline

Parameter & \multicolumn{2}{c}{Value} \\

\hline

$a$        & $0$ & $575073$ \\

$b$        & $0$ & $0383567$ \\

$c$        & $-0$ & $00144917$  \\

$\gamma_0$    &  $-0$ & $138971$  \\ 

$\gamma_1$    &  $-0$ & $0633399$  \\

$\beta_1^0$   &  $1$ & $04452$  \\

$\beta_2^0$   &  $0$ & $311702$  \\

$\beta_1^1$   &  $0$ & $872563$  \\

$\beta_2^1$   &  $0$ & $225783$  \\

\hline \hline
\end{tabular}
\end{center}
\end{table}

\subsection{Fits to PCFs \label{subsec:Fits}}

\begin{figure}
\begin{center}
\includegraphics[clip,width=0.45\textwidth]{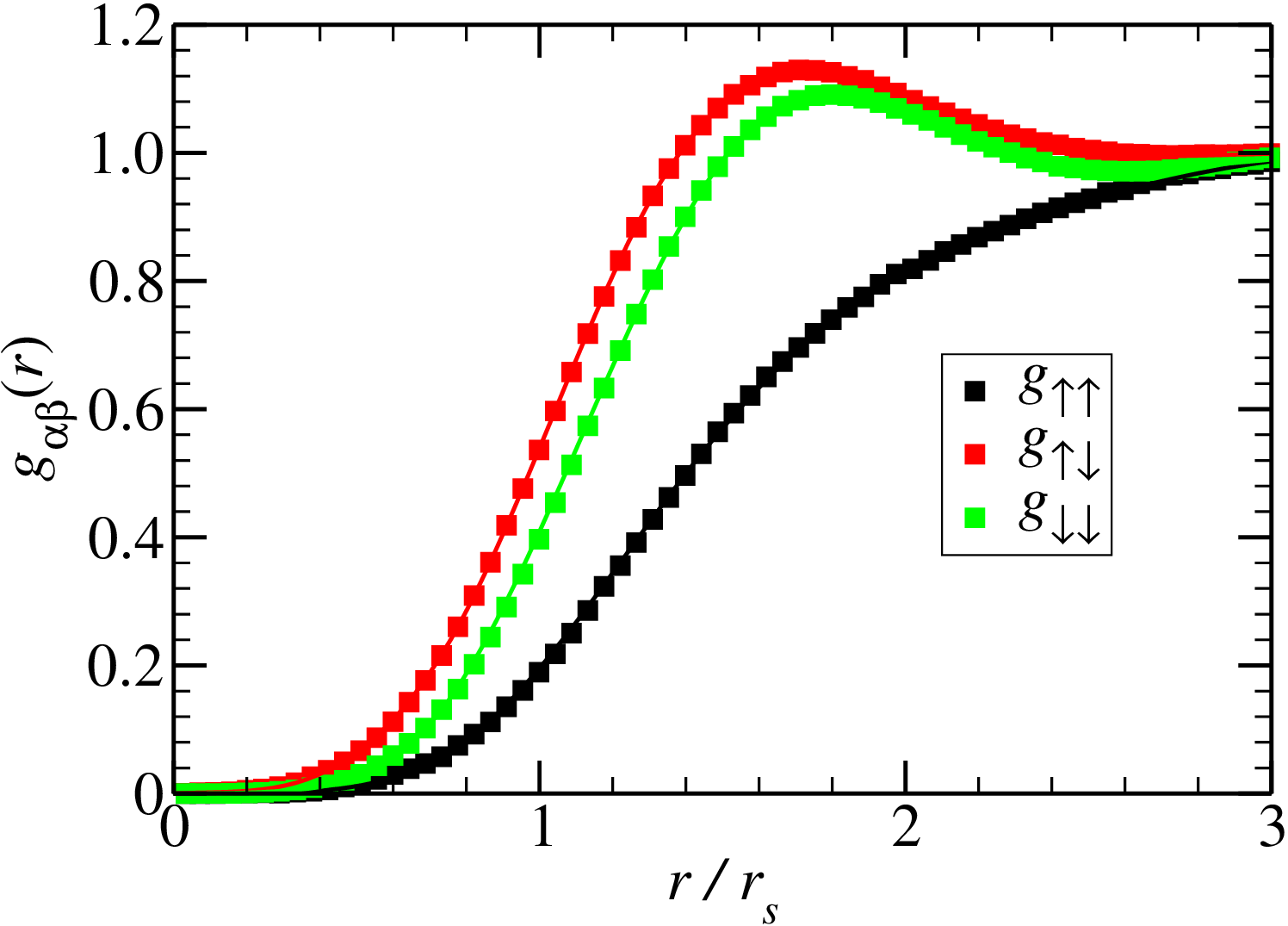}
\caption{(Color online) Cubic-spline fits to our twist-averaged extrapolated
  pair correlation function data at $r_\text{s}=20$ a.u.\ and $\zeta=0.66$. 
  Ten knots were used in the spline fits.  Square symbols represent the raw 
  QMC data points and lines represent our spline fits to those data.  In all 
  other figures, we have shown the raw QMC data rather than the spline fits 
  displayed here.
\label{fig:fits}}
\end{center}
\end{figure}

We have performed cubic-spline fits to our PCF data.  The spin
resolution across the density range $r_\text{s}=0.5$--20 a.u.\ and spin
polarizations $\zeta=0$, 0.34, 0.66, and 1 was well represented by
cubic-spline fits with ten knots for each spin-resolved PCF\@.  The boundary 
conditions used were those of natural splines: second derivatives set to 
zero at the first and last data points.  An example of
the fits obtained is given in Fig.\ \ref{fig:fits}, where we plot raw
QMC data together with fits to the data.  The splines reproduce our
estimate for the on-top pair density at $r=0$, and are accurate for
$0\leq r/r_\text{s} \leq 3$.  For unpolarized and fully polarized
systems, we fit the DMC data, whereas for partially polarized systems,
we provide fits to the extrapolated estimates of the PCFs.  Note that
we used down spins as the majority spins in our simulations.  
We did not impose any exact results as constraints when performing the 
spline fits, in order to give the most accurate fit possible to our raw data.  
As discussed above, our QMC data are consistent with the Kimball cusp 
conditions, and the spline fits follow the data closely in all systems 
studied.  On the other hand, our PCF data inevitably suffer finite-size 
effects at large-$r$, so various sum rules requiring integration of the PCF 
over all space will not be satisfied exactly.

These spline fits are available in the form of a small Fortran 90
program that we have written and which is included as part of the
Supplemental Material accompanying this article.\cite{supplemental}
Subroutines from the SLATEC Common Mathematical Library\cite{SLATEC}
were used to write this program.  SSFs can be obtained from these data
by Fourier transformation.

We also compared fitting schemes previously used to represent PCF data
for these systems, including those proposed by Ortiz and
Ballone,\cite{OB94} and Gori-Giorgi, Sacchetti, and Bachelet.\cite{GSB}
The former scheme struggled to describe the region around the
peak in the PCF with quantitative accuracy.  The latter scheme was
successful at high density (for $\zeta=0$), but less reliable
otherwise.\cite{GSB}  

\section{Conclusions \label{sec:conclusions}}

We have investigated ground-state properties of the 3D HEG using the VMC and DMC
methods.  Our simulations
cover the density range $r_\text{s}=0.5$--20 a.u.\ and spin
polarizations $\zeta=0$, 0.34, 0.66, and 1.  We used highly
accurate wave functions incorporating backflow and three-body correlations.
Twist averaging was used to reduce finite-size effects, and we show that 
this can be performed efficiently using a control variate method.

The spin resolution of the PCF and SSF are reported at each spin polarization
studied.  At $\zeta=0$, we obtain good agreement with previous QMC studies,
except for $r/r_\text{s} \lesssim 0.5$, where our data are significantly 
smaller than those of Gori-Giorgi \emph{et al}.\cite{GSB}  
The effects of twist averaging on these quantities are shown.  We
report the $r_\text{s}$- and $\zeta$-dependence of the energies for
118-electron HEGs.  It is hoped that the higher accuracy and lower noise than
earlier QMC studies that we have achieved, particularly of the
PCFs, will be valuable in, for example, aiding the construction of nonlocal
density functionals.  A small Fortran 90 program is available to reproduce
spline fits to our PCF and SSF data.

\begin{acknowledgments}
  The authors acknowledge financial support from the EPSRC\@.  
  Computational resources were provided by Cambridge University's High
  Performance Computing Service. We thank P.\ L\'opez R\'\i os and A.\ 
  J.\ Morris for useful discussions.
\end{acknowledgments}

\end{document}